\documentstyle[aps,prl,multicol,fancyheadings,epsfig,amsbsy,amssymb,amstex]{revtex}

\def\bbbc{{\mathchoice {\setbox0=\hbox{$\displaystyle\rm C$}\hbox{\hbox
to0pt{\kern0.4\wd0\vrule height0.9\ht0\hss}\box0}}
{\setbox0=\hbox{$\textstyle\rm C$}\hbox{\hbox
to0pt{\kern0.4\wd0\vrule height0.9\ht0\hss}\box0}}
{\setbox0=\hbox{$\scriptstyle\rm C$}\hbox{\hbox
to0pt{\kern0.4\wd0\vrule height0.9\ht0\hss}\box0}}
{\setbox0=\hbox{$\scriptscriptstyle\rm C$}\hbox{\hbox
to0pt{\kern0.4\wd0\vrule height0.9\ht0\hss}\box0}}}}
\newcommand{\beq}{\begin{eqnarray}} 
\newcommand{\eeq}{\end{eqnarray}} 

\begin{document}
\title{Impurity  induced resonant state in a pseudogap state of 
a high temperature superconductor.} 
\author{H.V. Kruis$^{1,2}$, I. Martin$^2$ and A.V. Balatsky$^2$}
\address{$^1$ Institute Lorentz, Leiden University, P.O. Box 9506,
 NL-2300 RA Leiden, The Netherlands, \\
 $^2$Theoretical Division, Los Alamos National
Laboratory, Los Alamos, NM 87545}

\date{Printed \today }

\maketitle

\begin{abstract}
We predict a resonance impurity state generated by 
the substitution of one $Cu$ atom with a nonmagnetic atom, such as $Zn$,
 in the {\em pseudogap} 
state of a high-$T_c$ superconductor.  The precise microscopic origin 
of the pseudogap is not important for 
this state to be formed, in particular this resonance will be present even in the absence of 
superconducting fluctuations in the normal state. 
In the presence of superconducting fluctuations, we predict the existence 
of a counterpart impurity peak on a symmetric bias.
 The nature of impurity resonance is similar to 
the previously studied resonance in the d-wave superconducting state.  

\end{abstract}

\pacs{Pacs Numbers: check Martin balatsky and balatsky salkola rosengren paper} 

\vspace*{-0.4cm}
\begin{multicols}{2}

\columnseprule 0pt

\narrowtext
\vspace*{-0.5cm}

The effects of a single magnetic and nonmagnetic impurity in high temperature 
superconductors have been studied intensively both theoretically
\cite{Balatsky,Flatte,Atkinson,Tanaka,Sachdev} and more recently experimentally by
scanning tunneling microscopy (STM)  \cite{Pan,Hudson,Yaz}. 
Understanding of the impurity states in high-$T_c$ materials is  important
because impurity atoms qualitatively
modify  the superconducting properties, and   these impurity induced
changes can be  
 used to identify the nature of the pairing state in superconductors. 

Up to date   theoretical analysis  of the impurity states has been focused
on the low temperature regime $T \ll T_c$ well below the superconducting
transition temperature $T_c$. On the other hand it is well known that in 
the normal state $(T \geq T_c)$ of underdoped cuprates, 
the electronic states at the Fermi energy are 
depleted due to pseudogap (PG) $\Delta_{PG}$, as was seen by STM 
\cite{Fischer} and by angular resolved photoemission \cite{ShenCampuzano}. 
One can  consider the temperature evolution of the impurity state as the 
temperature increases and eventually becomes larger than $T_c$.
 There are two possibilities for  the  evolution of impurity 
 resonance at $T>T_c$: a) the impurity resonance 
 gradually broadens until the superconducting gap
  vanishes at which point the impurity resonance totally disappears and b) 
the resonance gets broader however survives above $T_c$. 
Which of the possibilities is realized
 depends on the normal state phase into which the
superconductor evolves. It 
has been argued  \cite{Loram,Yanson} that in the underdoped regime the superconducting gap opens 
up in addition to the  pseudogap present well above $T_c$.  We find that the impurity resonance 
survives above $T_c$ in the {\em pseudogap} state of high-$T_c$ materials.
 The position and the 
width of the resonance are determined by the impurity scattering strength and PG scale.
In the absence of a PG above $T_c$ the impurity state disappears.

The origin of PG state is one of the most strongly debated issues. 
 Some models attribute PG to superconducting phase fluctuations 
 above $T_c$ \cite{kivelson95}; others to a competing non-superconducting 
 order parameter \cite{laughlin2000}.  Another possibility is that even
  within the PG regime there are at least two distinct
   sub-regimes -- $strong$ and $weak$
pseudogaps, with weak pseudogap occurring at higher temperatures
 due to antiferromagnetic fluctuations, and strong pseudogap being 
 related to superconducting fluctuations\cite{pines,ivar_stripes}.

\begin{figure}[htbp]
  \begin{center}
    \includegraphics[width = 3.0 in]{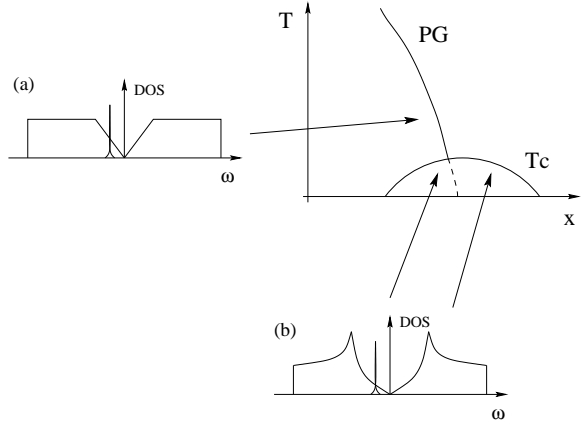}

\vspace{0.5cm}

\caption{An impurity state in a high $T_c$ superconductor:
(a) The DOS in the pseudogap regime used in this article (see also $\left[ 11 \right]$)
and (b) 
the DOS in the superconducting state as was used in $\left[ 1 \right]$. 
In both phases there is a resonant state.}
\label{phasediagram}
\end{center}
\end{figure}

In this article we  address the impurity induced resonance or quasibound 
state that is generated by a strong nonmagnetic impurity scattering in a $CuO$ 
plane in the {\em normal state} of high-$T_c$ materials. Specifically we 
calculate the resonant state generated by 
the substitution of one $Cu$ atom with a $Zn$ atom using  the 
self consistent $T$ matrix approach. We rely on the fact that the density of states (DOS) is 
depleted at the Fermi energy in the PG regime. We argue that the mere fact that DOS is 
depleted at the Fermi energy is sufficient  to produce   impurity 
resonance near the nonmagnetic impurity, such as $Zn$.  However  {\em  no 
particular use of the  superconducting correlations   above $T_c$  has been 
made} in our analysis. For example,  the results we present, will be valid in 
the PG state {\em with no superconducting  phase or amplitude fluctuations 
above $T_c$}, as long as there are interactions that lead to the PG state, 
as indicated  by a depleted DOS.  This is an important caveat that broadens 
the validity of the model regardless of the microscopic origin of the PG in 
the high-$T_c$ superconductor.
 The approach  we take is similar to the previous analysis of 
the nonmagnetic impurity in the superconducting state \cite{Balatsky}. 
See also figure \ref{phasediagram}.

The superconducting fluctuations are not required for the 
formation of the  impurity state in the PG regime.  However,
 in the presence of superconducting fluctuations an additional
  important feature of the impurity state is expected to appear. 
   The natural quasiparticles in the superconducting state are
    the Bogoliubov quasiparticles, which are a linear combination 
    of a particle and a hole.  As a consequence, an impurity 
    state in a superconductor appears both on  positive (particle) 
    and negative (hole) biases \cite{Balatsky,yazdani}.
In the phase-fluctuating pseudogap the Bogoliubov quasiparticles 
acquire a finite lifetime.  However, the particle-hole 
symmetry of the impurity states should remain to the extent 
to which the superconducting quasiparticles are defined 
\cite{Flatte}.  Hence, the PG impurity states can serve 
as an extremely local probe able to distinguish between 
the superconducting and non-superconducting scenarios
 for the PG regime.

To be specific we need a model DOS that captures the main features of the PG 
in high-$T_c$ materials. For this purpose we use the  DOS that    was 
measured  \cite{Loram} by  Loram {\it et al.}. In this work it has been 
argued that the DOS is a linearly vanishing function of energy  within the 
$\Delta_{PG}$ energy range near the Fermi surface, see figure 1(a). 

We find that   
 such a model indeed  gives rise to 
an impurity  bound state  with energy $\Omega'$ and decay rate $\Omega''$ 
equal to 
\begin{eqnarray}
 \label{solution}
 \Omega&=& \Omega' + i \Omega'' 
 \nonumber \\
 &=& - \frac{\Delta_{PG}}{2 U N_0} \frac{1}{ \ln |2 U
N_0|}
\left[ 
1 + \frac{ i \pi ~\mbox{sign} (U)}{2 \ln |2 U N_0|}
\right],
\end{eqnarray}
where we have assumed the impurity scattering  to be strong enough
so that the result can be calculated to  logarithmic accuracy
with $\ln |2 U N_0| > 1$ \cite{CommentB}. This is the main result of our 
work, which we will derive in the remainder of this paper.

The Hamiltonian for this single impurity problem is given by
\begin{equation}
H_{int} = U \hat{n}_0 = U \sum_{k k' \sigma} c^\dagger_{k\sigma}
c_{k'\sigma},
\end{equation}
where $U$ is the strength of the scalar impurity potential taken to be nonzero only at $r=0$. 
The scattering matrix  \cite{Balatsky} can be written as 
\begin{equation} 
T = \frac{U}{1 - U \sum_{\bf k} G_{\bf k} (\omega)} = \frac{U}{1 - U  G_0 
(\omega)}, 
\end{equation} 
with $G_0(\omega)$ the on site Green's function \cite{comment1}.
The states generated by the impurity
are given by the poles of the $T$ matrix: 
\begin{equation} 
\label{U} 
G_0 (\Omega) = \frac{1}{U}. 
\end{equation} 
This is an implicit equation for $\Omega$ as a function of $U$, the strength 
of the scattering. 
This solution can be complex, indicating the resonant 
nature of the virtual state. 
To solve this equation, we split $G_0$ into its imaginary and real part 
$G_0 = G_0' + i G_0 ''$. But also $G_0''(\omega) = - \pi N_0(\omega)$ with 
$N_0(\omega)$ the density of states. 

\begin{figure}[htbp]
  \begin{center}
    \includegraphics[width = 3.0 in]{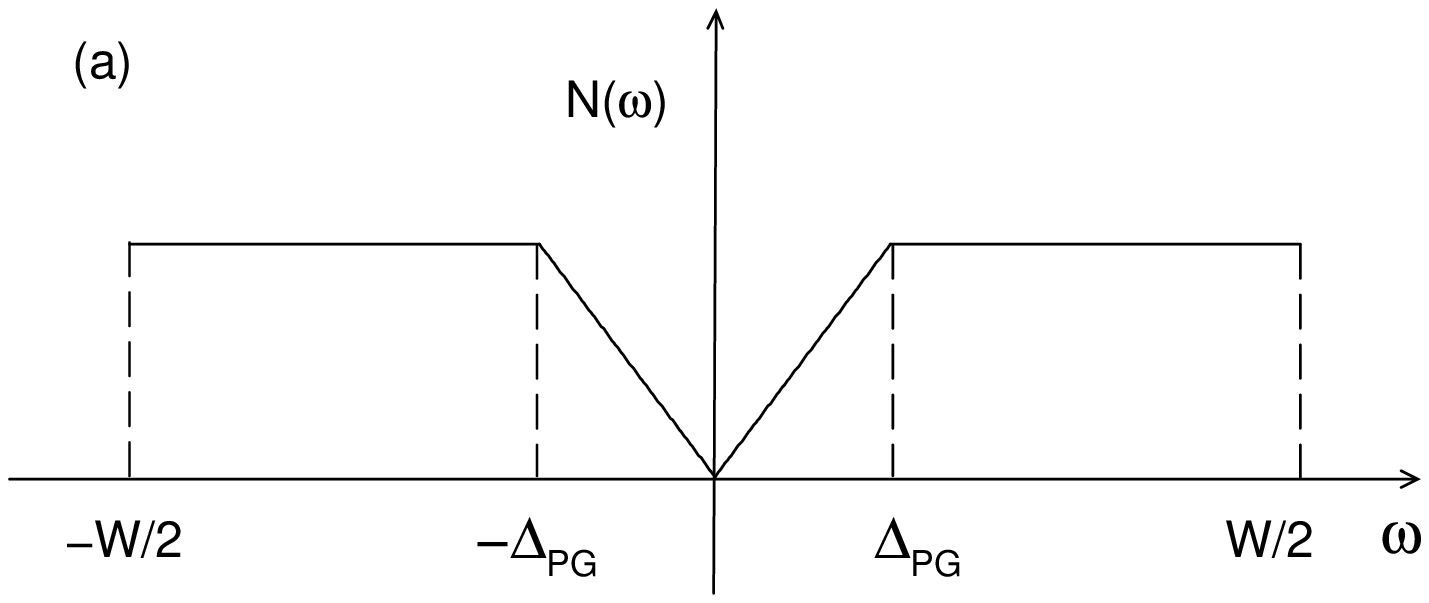}
 
   \vspace{0.25cm}
 
 \includegraphics[width = 3.0 in]{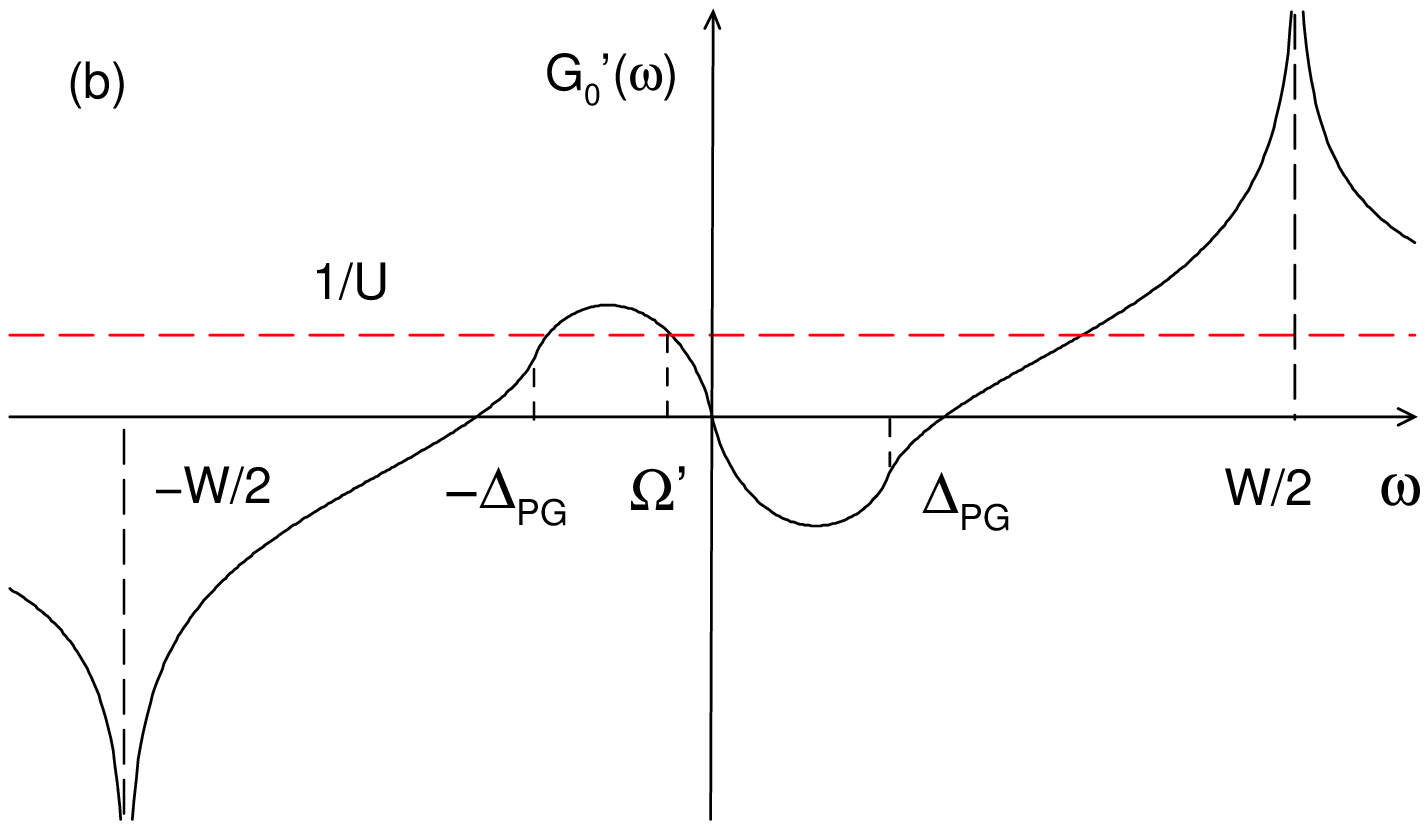}

\vspace{0.25cm}

 \includegraphics[width = 3.0 in]{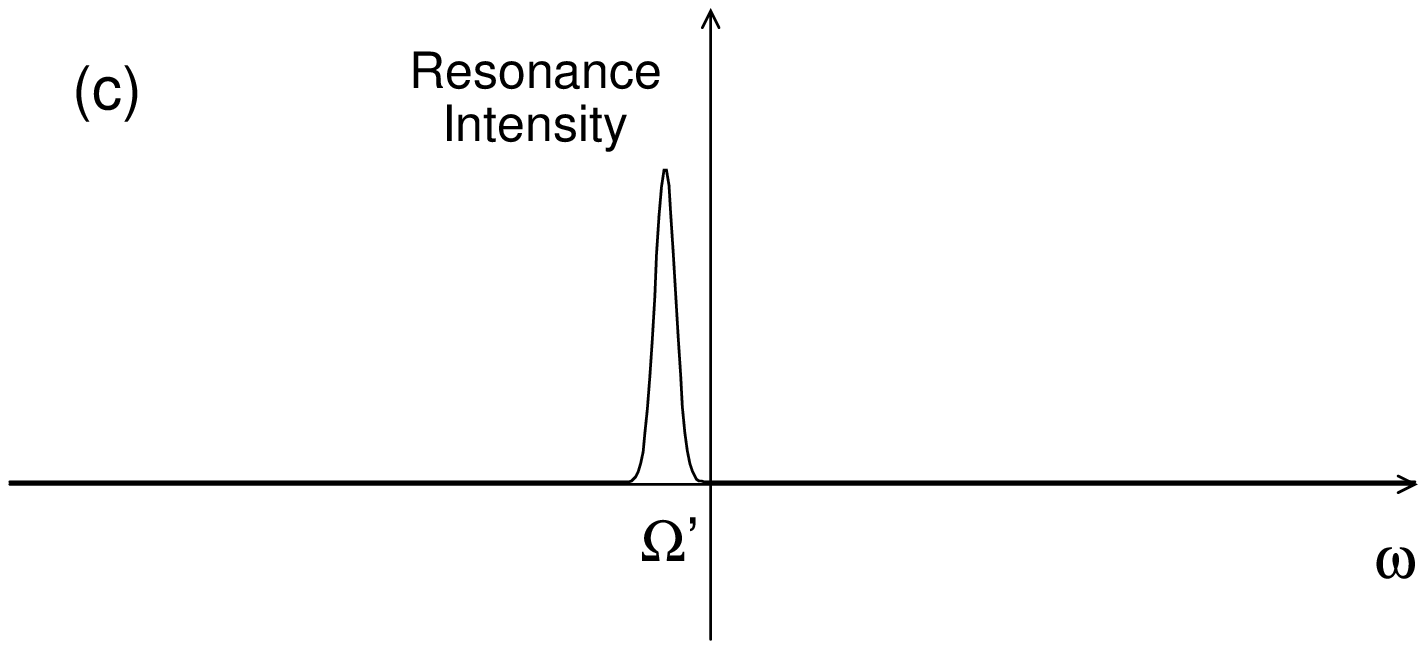}

\vspace{0.5cm}

\caption{(a) The density of states $N(\omega) = - G_0''(\omega) / \pi$.
Around the pseudogap 
states are only partly depleted e.g. $N(\omega) = N_0
|\omega|/\Delta_{PG}$,
where $N(\omega) = N_0$ for  $\Delta_{PG}< |\omega| < W/2$ with $W$
the
 bandwidth. (b) The real part $G_0'(\omega)$ of Green's function together
with $1/U$ and $U$ positive. $\Omega'$ is the real part of the solution of the 
equation $G_0 (\Omega) = 1/U$ close to zero and therefore with sharp bandwidth.
 (c) The  impurity induced resonance
at $\Omega ' = - \Delta_{PG}/2 U N_0 \ln (2 U
N_0)$. Because the other three solutions of equation (\ref{U})
have much broader bandwidth, they are not depicted here.
All the figures are taken on the impurity site.}
\label{fig:density}
\end{center}
\end{figure}

Measurements on the electronic specific heat by Loram {\it et al.}
\cite{Loram}
show that the normal
state pseudo gap opens abruptly in the underdoped 
region below a hole doping equal to $p_{crit} \sim 0.19$ holes/$CuO_2$.
Inspired by these data, we will assume that around the pseudogap region,
states are
  partly depleted and the density of states is linear, that is
 $N(\omega) = N_0  |\omega| / \Delta_{PG}$ for $|\omega| \leq
\Delta_{PG}$
 and $N(\omega) = N_0$ for  $\Delta_{PG}< |\omega| < W/2$ with $W$
the
 bandwidth.
 This density of states is depicted in figure \ref{fig:density}(a).
 As it is obvious from
the solution of equation (\ref{U}), the precise position
and the width of the resonance will depend on the
specific form of the PG. We will use this linearly
vanishing PG DOS. Results for the other form of
$N(\omega)$ can be obtained in a similar fashion \cite{CommentC}.

From the Kramer-Kronig relation \cite{Mahan}
\begin{equation}
 G_0' (\omega) = \frac{1}{\pi} \int_{- \infty}^{\infty} d \omega' G_0
''(\omega')
P \left(\frac{1}{\omega' - \omega}\right),
\end{equation}
with $P$ Cauchy's principle value,
one can calculate the real part $G_0'$ giving
\begin{eqnarray}
\label{kramer}
G'_0(\omega)& =& - N_0 \ln \left| \frac{\frac{W}{2} -\omega}
{ \frac{W}{2} +\omega}\right| +N_0 \ln \left| \frac{\Delta_{PG} -\omega}
{ \Delta_{PG}+\omega}\right| \nonumber \\&-& N_0 \frac{\omega}{\Delta_{PG}} 
\ln \left| \frac{\Delta_{PG}^2 -\omega^2}
{ \omega^2 }\right|.
\end{eqnarray}
This function is depicted in figure \ref{fig:density}(b) together with
 $1/U$. 
If $2U N_0 > e$, with $e$ Euler's constant, one can see from this figure
that equation (\ref{U}) has four solutions. But because the width of a resonance
state is proportional to $|\Omega|$, the only state with sharp width is
the solution with $|\Omega|$ close to zero and we will only consider this 
solution.
After expansion in $\omega$ of equation (\ref{kramer}) 
we arrive at an expression for
this solution $\Omega$ of equation (\ref{U}):
\begin{equation}
G_0(\Omega) = - \frac{2 \Omega N_0}{\Delta_{PG}} \left[ 
\ln \left| \frac{\Delta_{PG}}{\Omega} \right| - \frac{i \pi~ \mbox{sign} (U)}{2}
\right] = \frac{1}{U},
\end{equation}
which is solved to logarithmic accuracy by
 expression
(\ref{solution}).
Using this formula, taking $N_0 = 1~ \mbox{state}/eV$,
$\Delta_{PG}\sim 300 K \sim 30
meV$ and  the scattering potential  $U \approx 4 eV,$
we estimate  $\Omega \sim 20 K \sim 2 meV$ as was
found by Loram {\it et al.} 
\cite{Loram} This energy is close to the $Zn$
resonance energy $\omega_0 = 16 K$, seen in the
superconducting state \cite{Pan}.

The solution of the impurity state deep in the
superconducting regime involves two aspects: the
energy position and the width of the resonance and
secondly, the real space shape of the impurity state.
We have discussed the energy of the impurity state 
above. Great advantage of the on-site impurity
solution for the localized potential $U$ is that only
on-site propagator $G_0(\omega)$ enters into
calculation. Hence the knowledge of the DOS was
sufficient to calculate the impurity state. On the
other hand, to calculate the real space image of 
impurity induced resonance, one  would require  more
detailed knowledge of the Green's functions in the PG
regime. 
Quite generally, one would expect for  a d-wave 
like PG with nearly nodal points  along the
$(\pm \pi/2, \pm \pi/2 )$ directions,
that the
impurity resonance in the pseudogap regime would be
four-fold symmetric, similar to
superconducting solutions
 \cite{Balatsky,Flatte,Atkinson,Tanaka,Sachdev,Pan,Hudson,Yaz}.
This calculation  would require a specific
model for the PG state and goes beyond the scope of this paper.

While no superconductivity is required to form the impurity
 state in the PG, if the superconducting fluctuations are 
 present then an additional satellite peak should appear 
 on a symmetric bias due to the particle-hole nature of 
 the Bogoliubov quasiparticles.  The relative magnitude
  of the particle and the hole parts of the impurity 
  spectrum can be used to determine the extent to which 
  the PG is governed by the superconducting fluctuations.  
  In the case of fully non-superconducting PG there should 
  be no observable counterpart state.  An optimal impurity 
  for such determination would appear to be $Ni$, which breaks 
  the particle-hole symmetry more weakly than $Zn$ even in the 
  superconducting state.
Combined with other experimental 
proposals \cite {janko,ivar_PG}, the impurity state can 
help to better understand the mysterious PG state.

In conclusion, we find  the resonance  state
that is induced by the nonmagnetic impurity  in the
normal state of a high-$T_c$ superconductor in the PG
regime where the DOS at the Fermi surface is depleted.
For the particular model of linear DOS we find that
impurity state energy is given by equation (\ref{solution}).
Impurity states survive at high temperature $T>T_c$
since the PG produces the DOS depletion. This
depletion is all that is necessary to produce the
intragap state. In our solution we did not rely on a
superconducting phase fluctuations above $T_c$ to
generate the impurity state. We estimated the energy
of the $Zn$ impurity resonance to be at $20 K$, assuming a
impurity potential equal to $U = 4 eV$. This resonance energy is
indeed close to the energy of the $Zn$ resonance at
$16K$ in superconducting state \cite{Pan}.

Acknowledgments: This work has been supported by US DOE. We are grateful
to J.W. Loram and S. Haas for useful discussions and to the Newton Institute, Cambridge,
where the idea of this work was developed. HVK acknowledges 
the hospitality offered by the Los Alamos National 
Laboratory during his three months
 stay.

%\vspace*{-2.2cm}

\end{multicols}


\begin{thebibliography}{99}
\bibitem{CommentB} The simplest model for thermal broadening is to assign the 
 temperature dependent width: Thermal broadening at high temperatures $T >T_c$
 substantially  broaden the impurity resonance peak $\Omega '' (T) 
 =\sqrt{(\Omega ''(T=0))^2 + T^2}$.
\bibitem{comment1} Even if there are any 
superconducting correlations in the d-wave pairing channel with an anomalous 
fluctuating propagator $F_{\bf k}(\omega)$,  it will not enter into the above 
equation. This is because regardless of the nature of the fluctuation we take 
$\sum_{\bf k}   F_{\bf k}(\omega) = 0$ because of the d-wave ($\sim 
\cos(2\theta)$, where $\theta$ is the planar angle along the Fermi surface) 
nature of the pairing correlations.
\bibitem{CommentC} We argue that the appearance of the intragap impurity
state is a robust feature of any depleted DOS around the Fermi surface.
We also considered the model DOS with $N(\omega) =
N_0 \left[ a +  (1-a) \omega^2 / \Delta_{PG}^2 \right]$
which leads essentially to similar results as a function of the
impurity strength with a resonant state at 
\mbox{$\Omega = -\Delta_{PG} (1 + i  \pi a N_0 U)/ (4 N_0 U(1-a -\Delta_{PG}/W))
 $} \mbox{$\approx  -\Delta_{PG} (1 + i  \pi a N_0 U)/ (4 N_0 U(1-a)) $} when 
 $\Delta_{PG}/W$ is small.
\end{thebibliography}
\end{document}